\documentclass[10pt,journal,final,compsoc]{IEEEtran}

\usepackage{subcaption}
\usepackage[table]{xcolor}
\usepackage{algorithmic}
\usepackage{eqparbox}
\usepackage{algorithm}

\usepackage{tikz}
\makeatletter
\let\@tabclassz =\@classz
\let\@tabclassiv =\@classiv
\makeatother
\definecolor{lightgray}{gray}{0.9}
\definecolor{lightblue}{rgb}{0.96,0.96,1.0}

\usepackage[cmex10]{amsmath}
\usepackage{nicefrac}
\usepackage{ragged2e} 
\usepackage[normalem]{ulem}
\usepackage{wrapfig}
\usepackage{amssymb}
\usepackage{bm}
\usepackage{listings}
\usepackage{multirow}
\usepackage{enumitem}

\usepackage{graphicx}
\graphicspath{{images/}}

\usepackage[switch]{lineno}

\usepackage{color}

\newcommand\Tstrut{\rule{0pt}{2.6ex}}         
\newcommand\Bstrut{\rule[-0.9ex]{0pt}{0pt}}   

\usepackage{hyperref}
\usepackage{xcolor}
\hypersetup{
    colorlinks,
    linkcolor={red!50!black},
    citecolor={blue!50!black},
    urlcolor={blue!80!black}
}

\begin{document}

\title{Design and Optimization of Conforming Lattice Structures}

\author{Jun Wu, Weiming Wang, Xifeng Gao
\IEEEcompsocitemizethanks{
\IEEEcompsocthanksitem
Jun Wu and Weiming Wang
are with the Department of Design Engineering,
Delft University of Technology, Delft, The Netherlands.\protect\\
\IEEEcompsocthanksitem Xifeng Gao is with the Department of Computer Science, Florida State University, US. \protect\\
\IEEEcompsocthanksitem Weiming Wang is also with the School of Mathematical Sciences, Dalian University of Technology, China. \protect\\
\IEEEcompsocthanksitem
Corresponding Author: Jun Wu, E-mail: j.wu-1@tudelft.nl
}
\thanks{}}

\markboth{Wu \MakeLowercase{\textit{et al.}}: Design and Optimization of Conforming Lattice Structures}
{Wu \MakeLowercase{\textit{et al.}}: Design and Optimization of Conforming Lattice Structures}

\IEEEtitleabstractindextext{%
\begin{justify}
\begin{abstract}
Inspired by natural cellular materials such as trabecular bone, lattice structures have been developed as a new type of lightweight material. In this paper we present a novel method to design lattice structures that conform with both the principal stress directions and the boundary of the optimized shape. Our method consists of two major steps: the first optimizes concurrently the shape (including its topology) and the distribution of orthotropic lattice materials inside the shape to maximize stiffness under application-specific external loads; the second takes the optimized configuration (i.e. locally-defined orientation, porosity, and anisotropy) of lattice materials from the previous step, and extracts a globally consistent lattice structure by field-aligned parameterization. Our approach is robust and works for both 2D planar and 3D volumetric domains. Numerical results and physical verifications demonstrate remarkable structural properties of conforming lattice structures generated by our method.
\end{abstract}
\end{justify}

}

\maketitle

\IEEEdisplaynontitleabstractindextext

%
\IEEEpeerreviewmaketitle

\begin{figure*}[t]
\centering
\includegraphics[width=0.92\textwidth]{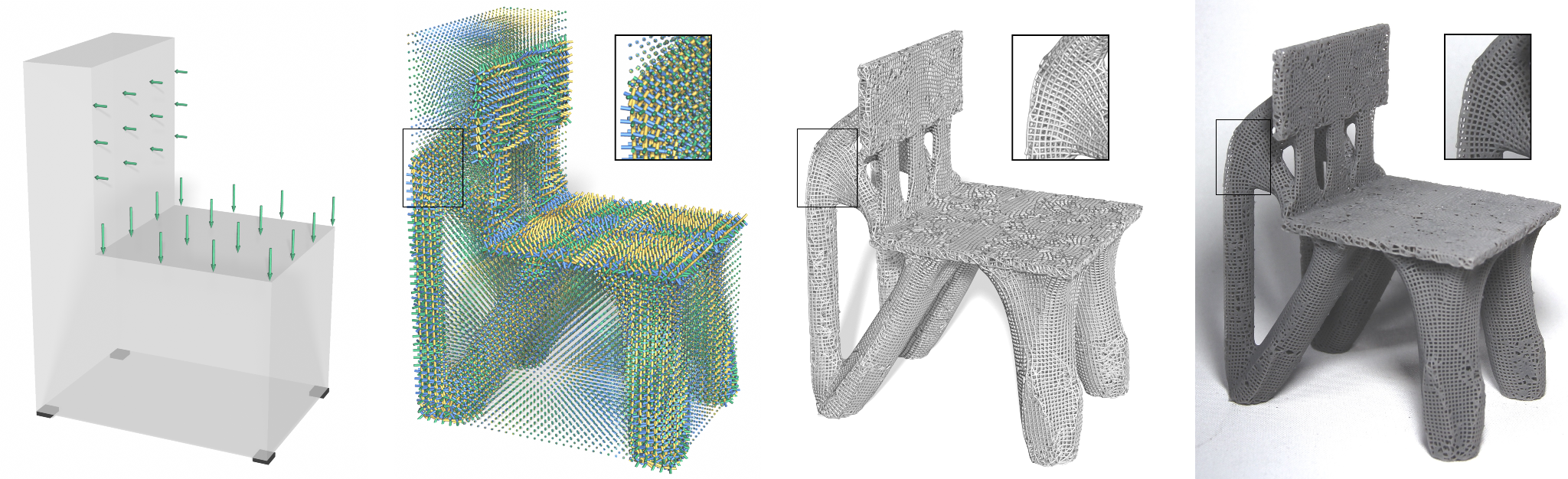}
	\caption{From left to right: Given a design domain with specified external loads, our method optimizes the distribution of lattice materials for maximizing stiffness. From the optimized, locally-defined lattice configuration, a globally connected lattice structure is compiled, and fabricated by 3D printing.
	}
	\label{fig:teaser}
\vspace{-4mm}
\end{figure*}

\section{Introduction}

The design of lightweight structures by optimization is a classical and still active topic in engineering. Stimulated by the increasingly high flexibility and resolution offered by 3D printing, there has been a growing interest in optimizing structures that are composed of delicate microstructures~\cite{Zhu2017TOG,Liu2018SMO}. These approaches assume that the microstructures are aligned with a prescribed regular grid. This simplifies modelling, simulation and optimization. It, however, limits the solution space, and thus the achievable structural performance. The microstructures are typically anisotropic (e.g. a hollowed cubic cell with uniform thickness is stiffer along its axes than along its diagonals). It is known that the orientation of anisotropic materials in stiffness-optimal structure coincides with the principal stress directions resulting from forces acting on these materials~\cite{Pedersen1989}. Furthermore, axis-aligned microstructures do not match the curved surfaces of 3D shapes. 

To address the above issues, in this paper we propose an efficient and robust method to generate \textit{conforming} lattice structures. A \textit{lattice} is a connected array of struts. The lattice structure generated by our method is conforming in two aspects: the struts align with principal stress directions, maximizing structural stiffness; and, struts on the boundary capture the curved surface of the optimized shape. We note that the shape, according to design options accessible to the user, is allowed to evolve together with the optimization of lattice distribution, i.e. the optimized shape is a subset of the design domain.

Our method has two major steps. In the first step, both the shape and the spatially-varying orientation of lattices inside the shape evolve simultaneously according to stress analysis and numerical optimization. Rather than relying on extremely high-resolution finite elements to capture the evolving lattice geometric details, we develop a homogenization-based topology optimization method which allows to efficiently simulate and optimize the lattice material distribution on a relative coarse level. By introducing a novel parameterization of the unit cell, our method ensures a uniform thickness of struts while allowing a sufficient degree in lattice anisotropy. The second step, which we call \textit{lattice compilation}, extracts a globally consistent lattice structure from the optimized, locally-defined lattice configuration, including orientation, porosity, and anisotropy. We address the challenging problem of extracting connected lattices across cells with spatially-varying orientation, by extending field-aligned meshing techniques. This extension allows a fast and robust lattice compilation where anisotropic geometric features are incorporated.

The specific contributions of our paper include:
\begin{itemize}
    \item A novel workflow for designing, in both 2D and 3D, conforming lattice structures based on homogenization-based topology optimization and field-aligned parameterization.
    \item Insights into optimal lattice structures from a mechanical perspective, analyzed via a detailed parameter study.
    \item A simple and effective parameterization of the unit cell for allowing structural anisotropy while ensuring a uniform thickness of struts.
    \item A new formulation to allow simultaneous optimization of the shape and the lattice distribution.
    \item A novel approach for extracting globally consistent lattice structures that accommodate anisotropy and heterogeneity.
\end{itemize}

Our method generates highly detailed lattice structures. The optimized lattice chair in Figure~\ref{fig:teaser}, for instance, consists of $178, 291$ struts, achieved on a simulation resolution of $140\times100\times200$. 

The remainder of the paper is organized as follows. In the next section we review related work. In Section~\ref{sec:overview} we give an overview of the proposed method. In Sections~\ref{sec:optimization} and~\ref{sec:comiplation} the two major steps of our method, lattice optimization and compilation, are presented. Results and analysis are presented in Section~\ref{sec:results}, before the conclusions are given in Section~\ref{sec:conclusions}.

\section{Related Work}
\label{sec:relatedWork}

\subsection{Structural Optimization for 3D Printing}

In the era of 3D printing (and more broadly, digital fabrication), structural optimization becomes increasingly relevant in computational design~\cite{Attene2018design}. Skin-frame structures~\cite{Wang2013TOG}, honeycomb-like Voronoi structures~\cite{Lu2014TOG}, tree-like supporting structures~\cite{Zhang2015CAGD}, and bone-inspired porous structures~\cite{Wu2018TVCG} have been optimized as lightweight infill for prescribed 3D shapes. Guided by outputs from structural optimization, Martinez et al. proposed to use graded orthotropic foams as a parameterized metamaterial to fill a prescribed shape~\cite{Martinez2016TOG,Martinez2017TOG}. In contrast to design and optimize internal structures for prescribed shapes, our method optimizes concurrently the shape and its internal microstructures for application-specific loads. Different from two-scale structural optimizations (e.g.~\cite{Xia2014CMAME,Zhu2017TOG}) which assume axis-aligned microstructures, our method optimizes the orientation of microstructures, in particular, to align it with spatially-varying stress directions. We restrict our design method to lightweight microstructures that are composed of struts, i.e. lattice structures.

Lattice structures are typically aligned with a regular grid~\cite{Dong2017JMD,Tamburrino2018JCISE}. Rosen and his co-authors proposed a method to design lattice structures that conform with the boundary surface of a prescribed 3D shape~\cite{Wang2005CIE,Nguyen2012SFF}. Our method optimizes concurrently the shape and align the lattices with stress directions. The alignment of structures along principal stress directions improves structural performance~\cite{Michell1904,Pedersen1989}. This principle has been applied to 2D planes (e.g.~\cite{Kwok2016CAD}) and curved surfaces~\cite{Tam2016stress,Kilian2017TOG,Li2017PG}. The appealing 2D results are achieved by tracing stress directions or based on a ground structure approach~\cite{Smith2002TOG,Zegard2014SMO}. Due to their inherent challenges associated with the initialization of samples/nodes, a direction extension of these methods to 3D volumetric lattices is not applicable. Our method constructs stress-aligned 3D volumetric lattices, relying on homogenization-based topology optimization and field-aligned meshing.

Our method is among recent efforts to structural analysis and optimization for 3D printing. Stava et al. proposed a method to detect and correct structural defects~\cite{Stava2012TOG}. Recent efforts include worst-case structural analysis~\cite{Zhou2013TOG,Panetta2017TOG}, and stochastic structural analysis~\cite{Langlois2016TOG}. Chen et al. proposed a solver for inverse elastic shape design~\cite{Chen2014TOG}. Ulu et al. optimized structures under force location uncertainty~\cite{Ulu2017TOG}. Our method, targeting on stiffness maximization of lattice structures under certain static loads, is complementary to these efforts. Yet the integration goes beyond the scope of this paper.

\subsection{Homogenization-based Topology Optimization}
\label{subsec:related_homo}
Topology optimization is an important design method for 3D printing, as it effectively leverages the fabrication flexibility to create structures with exceptional (mechanical) properties. Topology optimization transforms optimal shape design as a material distribution problem. In their seminal work, Bends{\o}e and Kikuchi proposed a homogenization method, which optimizes the distribution of square unit cells with variable rectangular holes~\cite{Bendsoe1988CMAME}. Due to the lack of manufacturing means for such microstructures back then, the homogenization method was replaced by density-based approaches (e.g. SIMP~\cite{Sigmund2001SMO}) 
which have since been widely used in industry and in many academic contributions (e.g. large scale optimization~\cite{Wu2016TVCG,Aage2017Nature,Liu2018TOG}). 

In light of the capability of 3D printing to fabricate microstructures, the homogenization method was recently revisited to design structures with manufacturable microstructures~\cite{Groen2018IJNME,Allaire2018CMA,Boddeti2018SciRep}, based on the rectangular hole model~\cite{Bendsoe1988CMAME}. A challenge is to compile a continuous structure from hollowed cells that are defined on a regular grid, and that, after optimization, have different orientations. To this end, a projection approach proposed by Pantz and Trabelsi~\cite{Pantz2008JCO} was improved to connect the orthotropic microstructures~\cite{Groen2018IJNME,Geoffroy20183d}. The output structure is represented by high-resolution pixels or voxels.

\begin{figure*}[!ht]
\centering
\def\svgwidth{\linewidth}
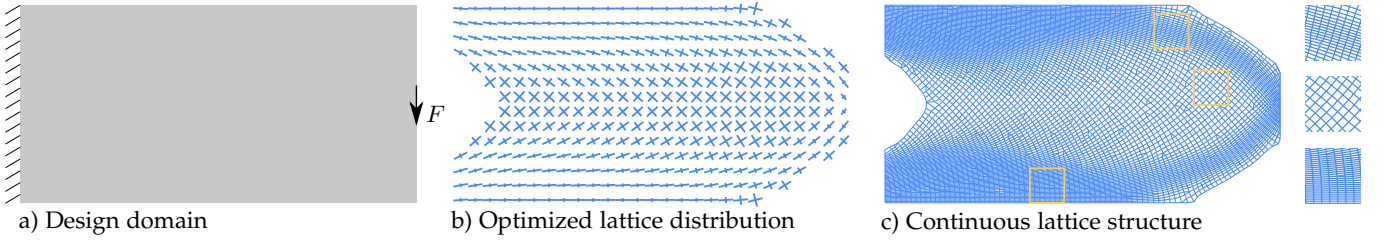
\caption{A 2D example, illustrating the pipeline of our approach. Given the design specification (a), the first step optimizes the distribution of lattice materials (b). The second step extracts a continuous lattice structure corresponding to the optimized lattice configuration (c).}
\label{fig:pipeline}
\vspace{-4mm}
\end{figure*}

Our approach to design conforming lattice structures follows the homogenization approaches, but differs in three aspects. First, we propose a new parameterization of cells to ensure that the variable cells have a constant thickness, while allowing a large degree in anisotropy. Uniform thickness is a common practice in (metal) additive manufacturing of (axis-aligned) lattices~\cite{Dong2017JMD,Tamburrino2018JCISE}. Such structures can also be fabricated by a direct extrusion in 3D~\cite{Mueller2014UIST}, or by robotic fabrication~\cite{Wu2016TOG,Huang2016TOG}. We note that variable thickness is not impossible with 3D printing. However, accurately achieving variable thickness requires precise control of the material flow rate in 3D printing hardware and robust algorithms in slicing~\cite{Kuipers2019}. Ensuring a uniform thickness alleviates these requirements. Second, our method simultaneously optimizes the lattice distribution and the shape, achieved using multiple design variables. Last but not least, while existing works exploited projection methods for generating high-resolution pixel or voxel models, we develop a novel approach based on field-aligned meshing to compile the lattice structure. The optimized structure is compactly represented by a graph. This direction shares a similar goal with the recent work by Arora et al.~\cite{Arora2018:arxiv}. In contrast to the design approach~\cite{Arora2018:arxiv}, our method unlocks a large solution space by optimizing the porosity, anisotropy, and orientation of lattices.

\subsection{Field-aligned Parameterization}

We develop a lattice compilation method based on field-aligned parameterization which has been researched intensively in the past decade, especially for generating quadrilateral (quad-) mesh. We review briefly the more recent works on hexahedral (hex-) meshing, and for quad-meshing we refer an interested reader to the survey by Bommes et al.~\cite{Bommes2013Survey}.

For a given 3D closed shape, field-aligned hex/hex-dominant meshing techniques typically consist of three steps~\cite{NieserSGP11,Huang2011,Li2012,Jiang2013,Sokolov2016TOG,Gao2017}. It starts by estimating the gradients of a volumetric parameterization using a directional field~\cite{Vaxman2016CGF,Ray2016}, where the field is discretized per vertex or per tetrahedron and smoothly interpolated within the volume under a boundary alignment constraint. This is followed by computing a parameterization aligned with the estimated gradients by fitting. Finally it extracts the hex-mesh in the parametric space~\cite{Lyon2016HexEx}. Robust hex-meshing remains a challenging problem. A promising direction is to topologically correct the directional fields~\cite{Li2012,Jiang2013,Solomon2017,Liu2018}. Lei et al. introduced a hex-mesh generation method based on surface foliation theory~\cite{Lei2017CMAME}. This approach, however, requires heavy topological pre-processing of the input.

The field-aligned parameterization pipeline is primarily used for generating semi-regular meshes. To ensure the validity of the mesh, complex geometric and topological computations are involved. In this paper we make use of field-aligned parameterization to generate lattice structures. This new application differs from mesh generation, as lattices are encoded by graphs rather than meshes. This goal sidesteps the numerical stability issue and geometrical and topological complexities typically occurred during mesh extraction from the parameterization.

To efficiently and robustly extract consistent lattice structures, we extend the robust meshing approach that was proposed by Jacob et al.~\cite{Jakob2015Instant} and further developed by Gao et al.~\cite{Gao2017}. The per vertex local parameterization from~\cite{Gao2017} fits our purpose well since the local parameterization aligns exactly with the direction field by construction and permits fast and scalable computations. The extension proposed in this paper allows to incorporate anisotropy and heterogeneity.

The recent work by Arora et al~\cite{Arora2018:arxiv} shares a similar goal as ours, i.e. to extract field aligned struts from stress directions. Our approach takes the optimized stress fields as input, without a field smoothing operation that compromises the accuracy of input fields. During lattice compilation, while they extract the struts by tracing stress lines and simplifying the duplicated ones, our approach directly generates struts by simple and efficient graph operations. This makes our approach fast and scalable, taking a couple of minutes for an input with tens of millions of tetrahedral elements (see Table~\ref{table:statistics}).
\section{Overview}
\label{sec:overview}

Given a design domain and application-specific loads, our method generates a lattice structure that maximizes structural stiffness. The struts in the optimized lattice structure conform with principal stress directions. Moreover, the struts on the boundary span a smooth surface faithfully approximating the optimized shape. 

As illustrated in Figure~\ref{fig:pipeline} for 2D and Figure~\ref{fig:teaser} for 3D, our approach consists of two steps. The first optimizes the shape (including its topology) and the distribution of lattice material within the shape. The input includes a design domain and boundary conditions (Figure~\ref{fig:teaser} left and Figure~\ref{fig:pipeline}a), as well as design specifications such as the target fraction of solid material. The design domain in 3D is represented by a closed triangle surface mesh. This mesh is voxelized, generating finite elements for simulation and optimization. The output is a set of fields, indicating, per element, the occupancy of lattice material, and the orientation and anisotropy of lattice material (Figure~\ref{fig:teaser} second left and Figure~\ref{fig:pipeline}b). A surface mesh is then reconstructed, representing the optimized shape, i.e. the interface between elements that are filled with lattice material and that are empty. The shape enclosed by this surface mesh is tetrahedralized. The optimized fields are then interpolated on the vertices of the tetrahedral model. The vertices, including their connectivity and their associated field values, serve as input for the second step, which compiles a globally connected lattice structure that are composed of struts (Figure~\ref{fig:teaser} second right and Figure~\ref{fig:pipeline}c). The output lattice structure is encoded by a graph.

\section{Lattice Optimization}
\label{sec:optimization}

\begin{figure}[tb]
\centering
\def\svgwidth{0.8\linewidth}
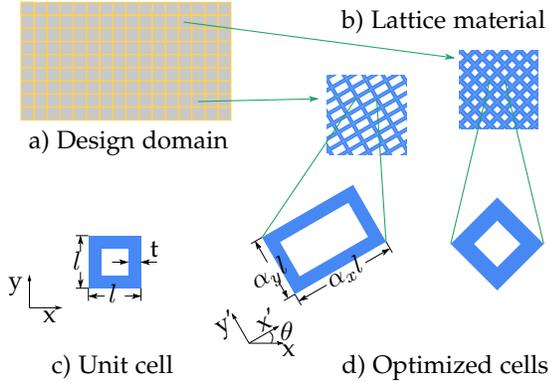
\caption{The design domain (a) is discretized into bilinear quadrilateral elements. Each element is filled with lattice material (b), i.e. a block of periodic cells (d). The cells are adapted by scaling and rotating a unit cell (c).}
\vspace{-4mm}
\label{fig:para}
\end{figure}

The goal of our optimization is to find the optimal distribution of lattice material that maximizes structural stiffness, subject to a number of design constraints. To this end, the design space is discretized into a regular grid of bilinear square elements in 2D or trilinear cubic elements in 3D. As illustrated in Figure~\ref{fig:para} for a 2D rectangular design domain, each element is to be filled by repeating a unique, rectangular-shaped cell. The cells are adapted from a unit cell by scaling and rotation. The scaling factors and rotation matrices are to be optimized. The scaling factors for the cell in element $e$ are denoted by $\alpha_e$, and in 2D by $(\alpha_{e,x}, \alpha_{e,y})$ and in 3D by $(\alpha_{e,x}, \alpha_{e,y}, \alpha_{e,z})$. The rotation matrix for element $e$ is denoted by $R_e$.

The unit cell is a hollowed square with a side length of $l$ and a thickness of $t$, which are specified by the user. To ensure a uniform thickness of struts across the design domain, during scaling the side length of cells is elongated, while the thickness ($t$) is kept constant. This creates cells with gradation in the fraction of solid material ($v_e$),
\begin{equation}
v_e(\alpha_{e}) =1 - \frac{(\alpha_{e,x} l-2t)(\alpha_{e,y} l-2t)}{\alpha_{e,x} \alpha_{e,y} l^2}.
\end{equation}
This gradation allows the optimization to place adapted cells with a smaller fraction of solid material in regions where the stress is relatively small. Furthermore, per axis elongation allows to increase the mechanical anisotropy of cells. This is beneficial since the stress tensors are typically anisotropic.

Besides a scaling factor per axis and a rotation matrix, each element is assigned a variable $\varphi_e$, to indicate whether the element is empty ($\varphi_e=0$) or filled ($\varphi_e=1$) with lattice material. The set of elements that are filled with lattice material defines the overall shape of the optimized structure. To allow for gradient-based numerical optimization, the variable $\varphi_e$ is relaxed to take intermediate values, i.e. $\varphi_e \in [0,1]$. This variable is akin to the density variable in classical density-based topology optimization, which in that context indicates the fraction of \textit{solid} material. In the context of lattice optimization, this variable shall be interpreted as the fraction of \textit{lattice} material. The fraction of solid material per element ($\rho_e$) depends on $\varphi_e$ and the fraction of solid material within an adapted cell ($v_e$), i.e.
\begin{equation}
    \rho_e (\varphi_e,\alpha_{e}) = \varphi_e v_e (\alpha_{e}).
\end{equation}

As the design space is parameterized by the fraction of lattice material ($\boldsymbol{\varphi}$), scaling factor ($\boldsymbol{\alpha}$), and orientation matrix ($\boldsymbol{R}$), the optimization problem is given as  
\begin{subequations}
  \begin{align}
    \underset{\boldsymbol{\varphi},\boldsymbol{\alpha},\boldsymbol{R}}{\min} &\quad  J  = \frac{1}{2} \mathbf{F}^{\mathsf{T}} \mathbf{U}(\boldsymbol{\varphi},\boldsymbol{\alpha},\boldsymbol{R}) 
    \label{eq:obj} \\
    s.t.: &\quad \textstyle{\sum_e \rho_e (\varphi_e,\alpha_{e}}) \le \overline{v}N \label{eq:volumeCondition}\\
     &\quad \varphi_e \in [0.0, \; 1.0],  \; \forall e \label{eq:densityCondition} \\
     &\quad \alpha_{e,k} \in [\underline{\alpha}_k, \; \overline{\alpha}_k],  \;  k\in\{x,y,z\}, \forall e . \label{eq:scalingCondition}
  \end{align}
  \label{eq:optFormulation}
\end{subequations}
Here the objective is to minimize the work done by the external force, which is equivalent to  minimize compliance (i.e. stiffness maximization). $\mathbf{F}$ denotes the force vector that is applied to the design domain. The force vector is constant. $\mathbf{U}$ denotes the displacement vector of the shape when it comes to its static equilibrium under the external force $\mathbf{F}$. The first constraint, Eq.~\ref{eq:volumeCondition},  restricts the amount of solid material, where $\overline{v}$ is the fraction of available solid material, and $N$ is the number of elements in the design domain. The second constraint, Eq.~\ref{eq:densityCondition}, sets bounds for the fraction of lattice material ($\boldsymbol{\varphi}$). The third constraint, Eq.~\ref{eq:scalingCondition}, sets bounds for the scaling factors ($\boldsymbol{\alpha_x}, \boldsymbol{\alpha_y},\boldsymbol{\alpha_z}$). The lower and upper bounds of the scaling factors are user-defined. 

The novelty of this formulation is two-folds. First, by optimizing the scaling factors rather than the thickness of hollowed cells, it ensures that all struts in the optimized structure have the same thickness. As discussed in Section~\ref{subsec:related_homo}, this eases the control of the 3D printing process. Second, we assign an additional variable $\varphi$ to indicate the occupation of lattice material. This makes the formulation more general. Prescribing $\varphi=1$ leads to optimized lattices that fill the entire design domain. This is useful as infill for prescribed shapes. Allowing $\varphi$ to be decided by the optimization enables both the shape and the lattice to evolve simultaneously, achieving a higher stiffness.


\subsection{Stiffness Matrix for Lattices}
\label{subsec:stiffness}
The objective function, Eq.~\ref{eq:obj}, involves the displacement vector ($\mathbf{U}$), which is related to the external force ($\mathbf{F}$). The unknown $\mathbf{U}$ is computed by solving the equilibrium equation with finite element analysis,
\begin{equation}
    \mathbf{K}(\boldsymbol{\varphi},\boldsymbol{\alpha},\boldsymbol{R}) \mathbf{U}=\mathbf{F}.
    \label{eq:elasticity}
\end{equation}
Here the stiffness matrix, $\mathbf{K}$, is assembled from per element stiffness matrix, $\mathbf{K}_e(\varphi_e, \alpha_e,R_e)$.

In standard finite element analysis of 
solids~\cite{Bathe2006finite}, the element stiffness matrix $\mathbf{K_e}$ is computed by integrating over the domain of an element, $\Omega_e$, 
\begin{equation}
\mathbf{K}_e = \int_{\Omega_e}{B^{\mathsf{T}}D_e Bdx},
\end{equation}
where $B$ is the element strain-displacement matrix for linear basis functions~\cite{Bathe2006finite}. $D_e$ is the fourth order elasticity tensor, computed based on the Young's modulus and Poisson's ration of the solid material.

For analyzing elements that are filled with lattice material, the elasticity tensor $D_e$ is not constant but rather depends on design variables $\alpha_e$ and $R_e$. Let us first consider an element that is filled lattice with $\varphi_e=1$. The stiffness matrix for lattices is calculated by 
\begin{equation}
    \mathbf{K}_e(1, \alpha_e,R_e) = \int_{\Omega_e}{B^{\mathsf{T}}D_e(\alpha_e,R_e)Bdx},
\end{equation}
The elasticity tensor of a rotated lattice cell, $D_e(\alpha_e,R_e)$, is computed by rotating the elasticity tensor of this cell in its local coordinate system, $D_e (\alpha_e)$. In engineering notation, $D_e$ is represented as a ${3\times3}$ matrix for 2D problems or a ${6\times6}$ matrix for 3D. The rotation of tensor is realized by 
\begin{equation}
    D_e(\alpha_e,R_e) = \overline{R}_e (R_e) D_e (\alpha_e) \overline{R}_e^{\mathsf{T}}  (R_e),
\end{equation}
where the tensor rotation matrix $\overline{R}$ is given in the Appendix.

The effective elasticity tensor of an elongated cell, $D_e (\alpha_e)$, is evaluated by numerical homogenization. We make use of the Matlab code provided in~\cite{Andreassen2014CMS} and ~\cite{Dong2019JEMT} for homogenization in 2D and 3D, respectively. Given the scaling factors, the domain of the elongated unit cell is discretized by square finite elements with linear basis functions. To avoid performing homogenization for every $\alpha_e$ during the optimization process, we pre-compute $D_e$ for regularly sampled $\alpha$ values. In 2D, we fit a surface for every non-zero entry in $D$ over the 2D domain of $[\underline{\alpha}_x, \overline{\alpha}_x] \times [\underline{\alpha}_y, \overline{\alpha}_y]$. In 3D we use trilinear interpolation.

For elements with $\varphi_e$ between $0$ and $1$,
we use the power law from density-based approaches~\cite{Sigmund2001SMO} to interpolate,
\begin{equation}
    \mathbf{K}_e(\varphi_e,\alpha_e,R_e) = \varphi^p_e \mathbf{K}_e(1,\alpha_e,R_e),
\end{equation}
where the parameter $p$ (typically $p=3$) is introduced to penalize intermediate values in $\varphi_e$, and consequently the optimization steers $\varphi_e$ towards either $0$ or $1$.

\subsection{Solving}

The optimization problem (Eq.~\ref{eq:optFormulation}) is solved in an iterative process, as in~\cite{Groen2018IJNME}. In each iteration the following computational steps are performed, until the maximum change in design variables is smaller than a threshold (or the maximum number of iterations is reached).

First, the equilibrium equation (Eq.~\ref{eq:elasticity}) is solved, obtaining the displacement vector, $\mathbf{U}$. From the element displacement vector ($\mathbf{U}_e$), strain ($\overline{\epsilon}_e$) and stress ($\overline{\sigma}_e$) per element, in engineering notation, are calculated by
$\overline{\epsilon}_e = {B} \mathbf{U}_e$ and $\overline{\sigma}_e = D_e(\alpha_e,R_e) \epsilon_e$, respectively.

\begin{figure*}[t]
\centering
\def\svgwidth{\linewidth}
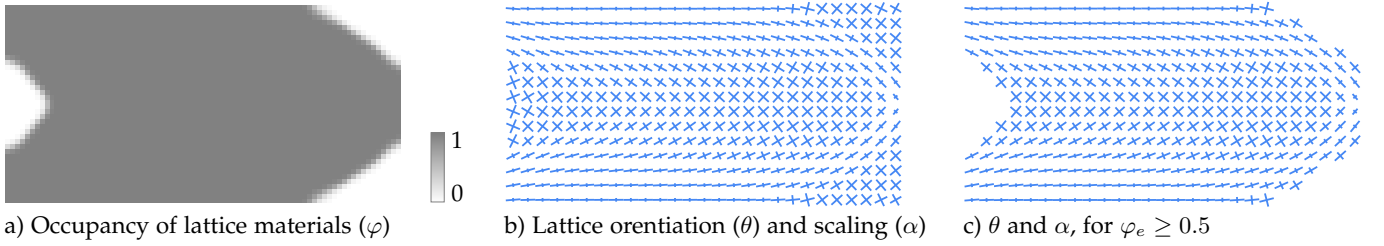
\caption{Visualization of the optimized fields.}
\label{fig:cantilever31}
\vspace{-4mm}
\end{figure*}

Second, design variables $\varphi$ and $\alpha$ are updated using a gradient-based solver. We make use of the method of moving asymptotes (MMA)~\cite{Svanberg1987IJNME}. To avoid checkerboard patterns, the design variables are regulated into $\tilde{\varphi}$ and $\tilde{\alpha}$ using the so-called density filter. $\tilde{\varphi}$ is then projected into $\overline{\tilde{\varphi}}$ by a (smoothed) Heaviside operation, to approach a 0-1 solution. The filter and Heaviside operator are widely used in density-based approaches, e.g. in e.g.~\cite{Wang2011SMO,Wu2018TVCG}.

Third, the orientation of each element ($R_e$) is updated based on the associated stress tensor ($\sigma_e$). The stress tensor is symmetric positive-definite. By eigendecomposition we obtain three mutually orthogonal principal stress directions ($v_1, v_2, v_3$). The eigenvectors are ordered by respective eigenvalues in ascending order, i.e. $\gamma_1 \le \gamma_2 \le \gamma_3$. As shown by Pedersen~\cite{Pedersen1989}, the optimal orientation of an orthotropic material coincides with the principal stress directions, hence the element is rotated by $R_e=[v_1^{\mathsf{T}}; v_2^{\mathsf{T}}; v_3^{\mathsf{T}}]$.

Fourth, the stiffness matrices of lattices, $K_e$, are re-calculated based on the updated orientation ($R_e$) and regulated variables $\overline{\tilde{\varphi}}$ and $\tilde{\alpha}$, according to Section~\ref{subsec:stiffness}.

\subsection{Example}
\label{subsec:example}

The output of our optimization is a set of fields defined on the design domain. Figure~\ref{fig:cantilever31} visualizes these fields for a rectangular domain, which is discretized by a grid of $80\times40$ elements. The unit cell has $l=10t$. The maximum fraction of solid material is $0.15$. Figure~\ref{fig:cantilever31}a shows the optimized lattice fraction field. Even with a fraction of solid material as small as $0.15$, the lattice covers a large portion of the design space. This is due to the fact that the unit cell has a small fraction of solid material (i.e. $36\%$, with $l=10t$). Figure~\ref{fig:cantilever31}b visualizes the orientation of optimized cells. Here the rotated frame is elongated according to the respective scaling factor per axis. For clarity the frames are shown for regularly-spaced samples. On the right, the frame field is visualized for elements which have a fraction of lattice material ($\varphi_e$) that is larger than a threshold ($0.5$).

\section{Lattice Compilation}
\label{sec:comiplation}
Up to this step, we have equipped with a design domain with a set of fields including fraction of lattice infill, orientation, and scaling, that are optimized for the prescribed external loads (cf Figure \ref{fig:cantilever31}). Since a region with a low fraction implies that few materials is required, we extract a sub-area (volume) from the design space by thresholding ($\ge0.5$) out low infill regions. With the actual shape being extracted, we now focus on generating a lattice structure that conforms to both the boundary of the shape and the directional and scaling fields. 

Our problem setting differs from the typical meshing problem in that both of our input and output are quite relaxed from the conditions of being a mesh. For the input, we put no constraints on its geometrical quality (i.e. angles, edge ratios, etc) nor its topological correctness (i.e. manifoldness, no holes, and free of intersections). This maximizes the scope of the problem but poses a great challenge to the design of a robust solution. For the lattice output, it does not require face (solid) elements, making complex topological operations in most of the meshing methods unnecessary for our purpose. Moreover, considering that it is not a hard requirement for our lattice structure to be all-hex cells for the designed structure to function, we choose the parameterization optimization in \cite{Gao2017} that can be easily adapted to handle graphs and propose a simple extraction strategy to generate a lattice structure. The produced lattice structure contains mostly quad (hex)-like connections while allowing certain irregularity to adapt for rapid changes in the directions and scales.

Our method takes a graph with vertices of the optimized shape as the input: $G = (V, E)$, where every vertex $\mathbf{v} \in V$ has a position $\mathbf{x} \in \mathbb{R}^k$ ($k$ is 2 for 2D and 3 for 3D), an orientation matrix $\mathbf{R} \in \mathbb{R}^{k\times k}$ encoding the cross directions and also denoting a local coordinate system, and a scaling vector $\mathbf{s} \in \mathbb{R}^{k}$ (i.e. $\boldsymbol{\alpha}$ in the previous section) composed of scales for the $k$ axes of the local coordinate system. Our goal is to extract a lattice structure, which is another graph $G' =(V', E')$ that (1) reproduces the input direction and the anisotropic property as much as possible, and (2) has a resolution that can be flexibly controlled by a target edge length $h$. 

In the following, we first describe the parameterization optimization that incorporates anisotropic orientations, and then present the lattice structure extraction.


\subsection{Parameterization}
Given an orientation field $\mathbf{O}$ that includes the cross directions for all the vertices, we want to compute a parameterization $\mathbf{P}$ with the gradient aligned to  $\mathbf{O}$. Methods that compute a global parameterization with the gradient aligning to the orientation field in a least-square sense (e.g. \cite{Bommes2009,Jiang:2015,Arora2018:arxiv}), rely on non-linear optimization solvers which are not scalable to large dataset. We instead compute a parameterization for the input graph by representing it with a set of local parameterization and minimizing an energy between the local parameterizations of adjacent vertices \cite{Jakob2015Instant,Gao2017}. 
\begin{wrapfigure}{r}{.5\columnwidth}
	\vspace{-10pt}
	\hspace{-10pt}
	\includegraphics[width=.5\columnwidth]{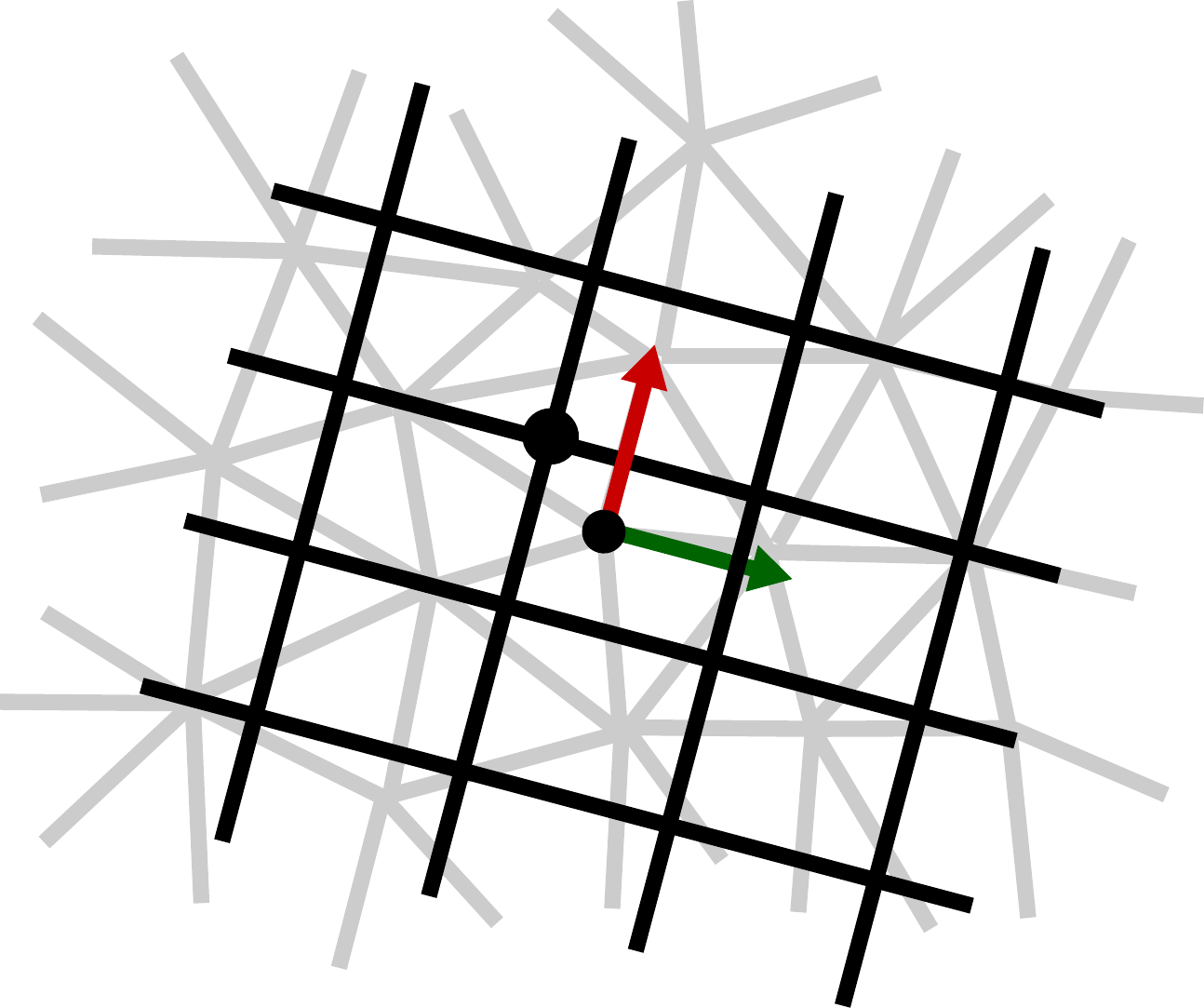}
	\put(-70,43){$\mathbf{x}$}
	\put(-78,65){$\mathbf{p}$}
	\label{fig:region}
	\vspace{-10pt}
\end{wrapfigure}The local nature of the parameterization makes it easily parallelizable and scalable to large inputs.

As illustrated in the inset, the local parameterization of a vertex in 2D plane (or 3D volume) can be uniquely determined by its origin $\mathbf{p}$, the orientation matrix $\mathbf{R}$, and unit lengths $h \cdotp \mathbf{s}$, where $h$ is the user-defined global target edge length. The unit lengths are fixed through the entire process. Unlike the previous approaches \cite{Gao2017,Jakob2015Instant} that treat directions as a 4 rotational symmetric field in 2D or a 24 rotational symmetric field in 3D, since the unit length varies for different axis, our coordinate system is mutable only by flipping the signs of each axis while maintaining the right-hand rule. The origin with a random initialization, is the variable we need to optimize.

Given the above setting, the optimization energy of the parameterization $\mathbf{P}$ is defined as the summation of all the squared differences of local parameterizations for each edge:

\begin{equation}
E(\mathbf{P}) = \sum_{i\in V}{\sum_{j \in N(i)}{||\mathbf{p}_i - (\mathbf{M}_{ij}\mathbf{t}_{ij} + \mathbf{p}_j)||^2}},
\end{equation}
where $N(i)$ is a set of all the vertices sharing an edge with vertex $i$,  $\mathbf{M}_{ij}$ is an interpolation of $\mathbf{M}_i$ and $\mathbf{M}_j$ where $\mathbf{M} = \mathbf{R}\mathbf{S}$ and $\mathbf{S}$ is the scaling matrix converted from $h\cdotp\mathbf{s}$, and $\mathbf{t}_{ij} \in \mathbb{Z}^k$ encodes the integer translations of $\mathbf{p}_j$. $\mathbf{M}_{ij}\mathbf{t}_{ij} + \mathbf{p}_j$ translates $\mathbf{p}_j$ by integer moves to the nearest position to $\mathbf{p}_i$, effectively avoiding the integer jumps between the two local parameterization and only the difference of their fractional parts is measured. The computation of $\mathbf{M}_{ij}$ is to interpolate the directions and scales separately,
\begin{equation}
\mathbf{M}_{ij} = \frac{(\mathbf{R}_i+\mathbf{R}_jr(\mathbf{R}_i,\mathbf{R}_j))}{||\mathbf{R}_i+\mathbf{R}_jr(\mathbf{R}_i,\mathbf{R}_j)||} \cdot \frac{(\mathbf{S}_i + \mathbf{S}_j)}{2},
\end{equation}
where $r(a, b)$ is the closest matching that gives the smallest difference between two coordinate systems which can be computed efficiently by enumerating all the cases. There are only two cases to compare in 2D and six cases for 3D.

The integer translation between two connecting vertices in the parameterization space, $\mathbf{t}_{ij}$, is computed by a rounding operation,
\begin{equation}
\mathbf{t}_{ij} = round [\mathbf{M}_{ij}^{-1}(\mathbf{p}_i - \mathbf{p}_j)].
\end{equation}
By doing so, the energy between the two vertices will be minimized. 
 
We minimize $E(\mathbf{P})$ in a Gauss-Seidel style by iteratively visiting every vertex and smoothing the origin of each vertex by computing an average of all its neighbors. The pseudo code of the optimization is provided in Algorithm~\ref{alg:optimizationRoutine}.
\begin{algorithm}
\caption{\label{alg:optimizationRoutine}Optimize-Parameterizations ($\mathbf{P}$)}
\begin{algorithmic}[1]
        \FOR{$i$ = 1, \ldots, $n$}
            \STATE $\mathbf{p}_i'$ $\gets$ $\mathbf{p}_i$, $d$ $\gets$ $0$
            \FORALL{$j \in \mathcal{N}(i)$}
                \STATE $\mathbf{p}_i'$ $\gets$ d$\mathbf{p}_i' + \mathbf{p}_j +\mathbf{M}_{ij}\mathbf{t}_{ij}$
                \STATE $d$ $\gets$ $d + 1$
                \STATE $\mathbf{p}_i'$ $\gets$ $\mathbf{p}_i'/d$
            \ENDFOR
            \STATE $\mathbf{p}_i$ $\gets$ $\mathbf{p}_i' + \mathbf{M}_{i} round\left[\mathbf{M}_{i}^{-1}(\mathbf{x}_i-\mathbf{p}_i')\right]$
        \ENDFOR
\end{algorithmic}
\end{algorithm}

\begin{figure}[h]
\centering
\includegraphics[width=\columnwidth]{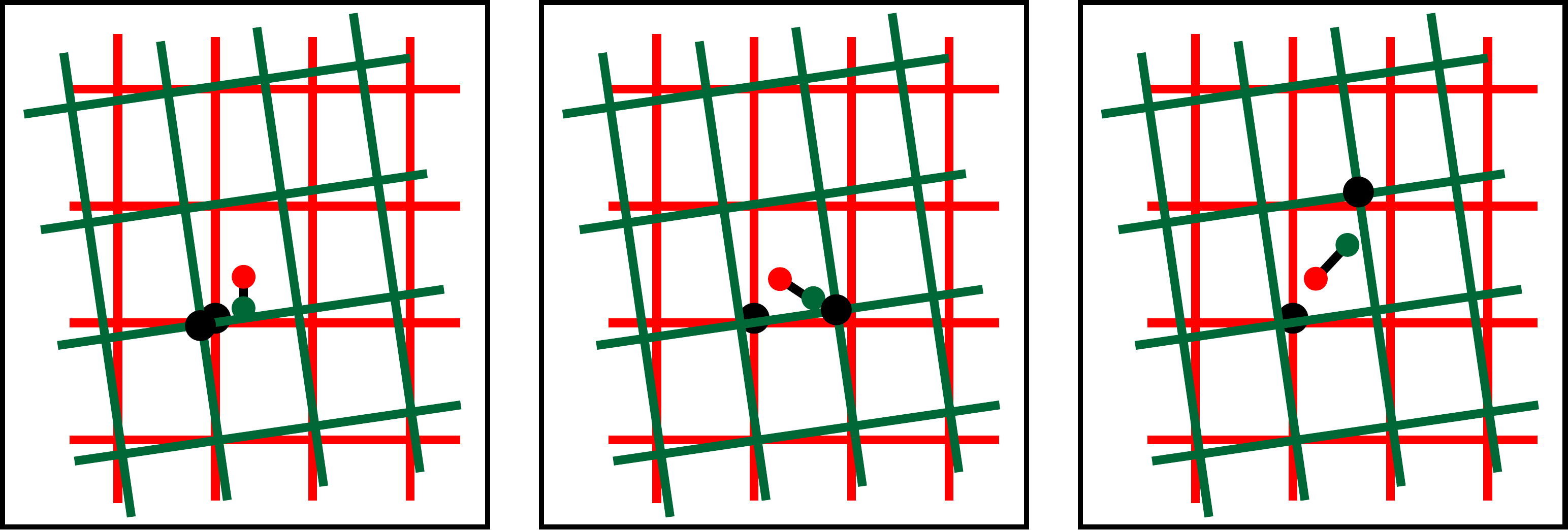}
\caption{Two close vertices in the input graph are not necessarily close in the parameterization space.}\label{fig:parameterization}
\end{figure}
The last step in line 8 rounds each origin of a local parameterization $\mathbf{p}_i$ to the integer position closest to the vertex position $\mathbf{x}_i$. Consequently, each component of $\mathbf{t}_{ij}$ becomes -1, 0, or 1. For example, as illustrated in Fig. \ref{fig:parameterization}, $\mathbf{t}_{ij} = (0,0)$ for Fig.~\ref{fig:parameterization} left, $\mathbf{t}_{ij} = (\pm 1,0)$ or $(0,\pm 1)$ for Fig.~\ref{fig:parameterization} middle, and $\mathbf{t}_{ij} = (\pm 1,\pm 1)$ for Fig.~\ref{fig:parameterization} right.

To speed up the optimization, similar to \cite{Gao2017}, we construct a hierarchical structure of the input graph by halving the number of vertices for each level and perform the optimization on each level of the hierarchy by 50 iterations for 2D and 200 iterations for 3D.

\subsection{Graph Extraction}

\begin{figure}[h]
\centering
\includegraphics[width=\columnwidth]{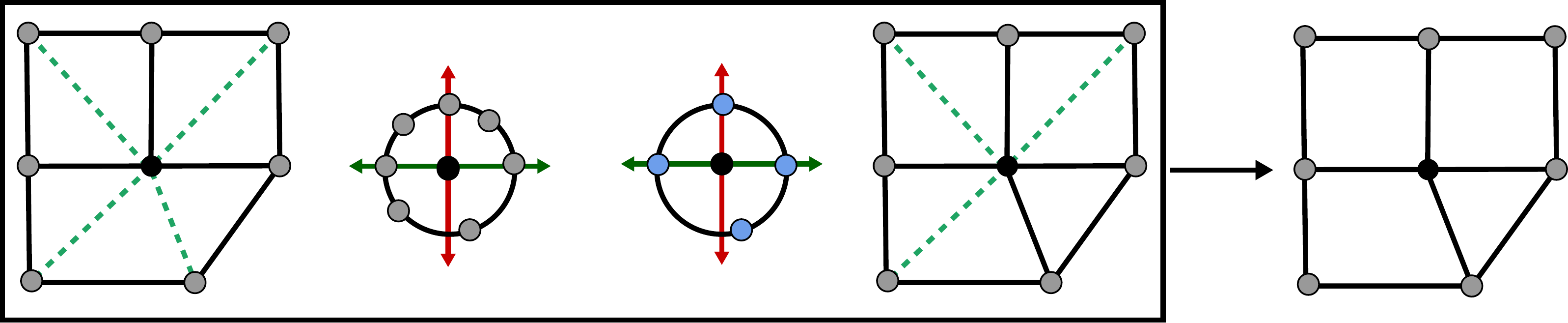}
\caption{Left: for a vertex in the graph, the nearest diagonal edges from its rotational directions will be relabelled to be maintained in the final graph if there is no edge representing the corresponding directions. Right: after the relabeling, our final graph is generated by discarding all the diagonal edges. }\label{fig:graph}
\end{figure}

In the input graph $G$, each vertex $\mathbf{v}$ has a smoothed local parameterization. The origin $\mathbf{p}$ of $\mathbf{v}$ provides a guidance for the vertex position in the output graph $G' = (V', E')$. Besides, the integer translation associated with each edge ($\mathbf{v}_i\mathbf{v}_j$) of $G$, $\mathbf{t}_{ij} \in \mathbb{Z}^k (k = 2, 3)$, categorizes this edge as a specific element in $G'$, depending on the $L_0$ norm of $\mathbf{t}_{ij}$ which is the number of $\pm 1$s in $\mathbf{t}_{ij}$. In 3D ($k=3$), this number can be 
\begin{itemize}
    \item 0 (i.e. $\mathbf{t}_{ij} = (0,0,0)$), indicating that the two vertices are very close in the parameterization space, and thus will be collapsed into to a point in $G'$,
    \item 1 ($\mathbf{t}_{ij} = (\pm 1,0,0), (0,\pm 1,0), \textrm{or}\, (0,0,\pm 1)$), meaning that the edge is parallel to one of the stress directions, and thus will be kept in $G'$,
    \item 2 ($\mathbf{t}_{ij} = (\pm 1,\pm 1,0), (\pm 1,0,\pm 1), \textrm{or}\, (0,\pm 1,\pm 1)$) or 3 ($\mathbf{t}_{ij} = (\pm 1,\pm 1,\pm 1)$), respectively corresponding to a rectangular or cuboid diagonal, which deviates from the stress directions and thus shall not appear in $G'$.
\end{itemize}
For example, black and dashed green edges in Figure~\ref{fig:graph} left correspond to $||\mathbf{t}_{ij}||_0 = $ 1 and 2, respectively.

By utilizing the positional guidance of $\mathbf{p}$ and the indication of $\mathbf{t}_{ij}$, the graph extraction is straightforward: collapse the edges with $||\mathbf{t}_{ij}||_0 = 0$ (dots in Figure \ref{fig:graph} represent the averaged positions of collapsed edges), keep the edges with $||\mathbf{t}_{ij}||_0 = 1$, and remove the diagonal edges (i.e. $||\mathbf{t}_{ij}||_0 = 2 \,\textrm{or}\, 3$).

While the above procedure generates a graph with mostly right angles, we notice T-junctions in the final graph with near flat angles that are suboptimal for the stiffness of the lattice structure. Figure~\ref{fig:graph} left illustrates a vertex with T-junction in 2D. This can be attributed to the fact that the removal of the diagonal edges is aggressive. The T-junctions appear near singularities of the parameterization (similar to the positional singularities in~\cite{Gao2017}) which result in elements with non-right angles, for example, triangles and pentagons in 2D, and prisms and general polyhedra in 3D. 

To address this issue, we propose to keep some diagonal edges in the final graph. Specifically, right after collapsing edges with $||\mathbf{t}_{ij}||_0 = 0$, we check the configuration of every vertex in the graph and identify critical diagonals. As illustrated in Figure~\ref{fig:graph}, for a vertex in black, the process is done by first normalizing all of its adjacent edge vectors onto a unit circle (sphere in 3D), then computing their nearest directions over 4 rotational-symmetric ones in 2D (6 in 3D), e.g. red and dark green arrows, and finally relabelling a diagonal edge to be an edge with $||\mathbf{t}_{ij}||_0 = 1$ such that each of the 4 (6 in 3D) stress directions is represented by an edge that is close to the direction (Figure \ref{fig:graph} right).

In summary, the process to extract the graph $G'$, i.e. a lattice structure, consists of the following steps.
\begin{enumerate}[leftmargin=*]
    \item Categorize the edges in $G$ based on $||\mathbf{t}_{ij}||_0$.
    \item Group vertices in $G$ according to $||\mathbf{t}_{ij}||_0$ such that groups are connected by edges with $||\mathbf{t}_{ij}||_0\neq 0$. Note that a group might contain only a single vertex.
    \item Generate the initial $G'$. For each group, a new vertex is positioned at the average of the origins of vertices in $G$. This vertex inherits the edges to new vertices that are converted from neighbouring groups.
    \item Categorize the edges in $G'$ based on $||\mathbf{t}_{ij}||_0$.
    \item Identify and relabel critical diagonals in $G'$ to avoid T-junctions, and remove other diagonal edges. 
\end{enumerate}

\begin{figure*}[t]
\centering
\includegraphics[width=\linewidth]{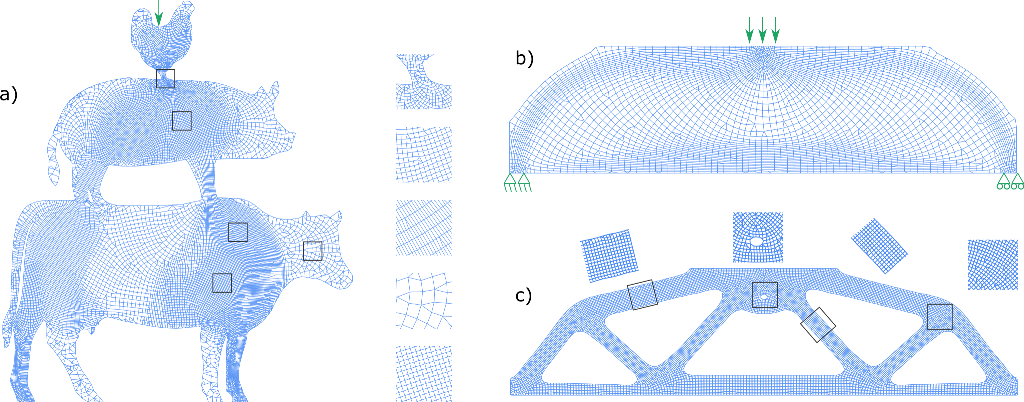}
\caption{Optimized 2D lattice structures for a prescribed freeform shape (a) and a rectangular design domain (b and c). The optimized lattice structures possess spatial variations in orientation, porosity, and anisotropy.
}\label{fig:lattice2D}
\vspace{2mm}
\centering
\includegraphics[width=\linewidth]{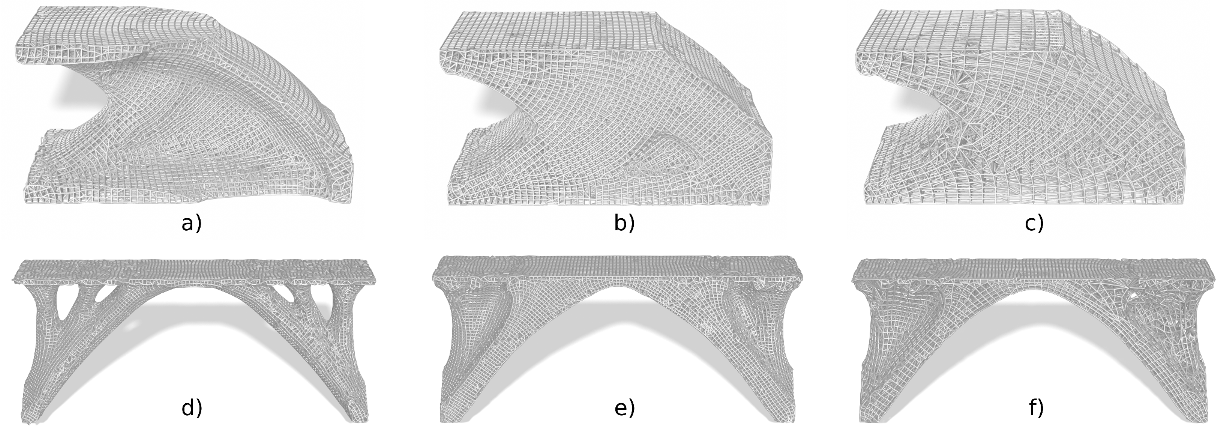}
\caption{3D lattice structures optimized from cuboid design domains, showing spatial variations in orientation, porosity and anisotropy. The design options are: (left) fixed $\alpha=1$ with design variables $R$ and $\varphi$, (middle) design variables $R$, $\varphi$, and $\alpha$ with $\alpha_x=\alpha_y=\alpha_z$, (right) full flexibility. With the increased design flexibility, the compliance reduces from left to right: 110.84 $\to$ 96.03 $\to$ 85.85 (cantilever), 230.52 $\to$ 177.86 $\to$ 149.96 (bridge).
}\label{fig:lattice3D}
\vspace{-4mm}
\end{figure*}

\begin{figure*}[tb]
\centering
\includegraphics[width=0.8\linewidth]{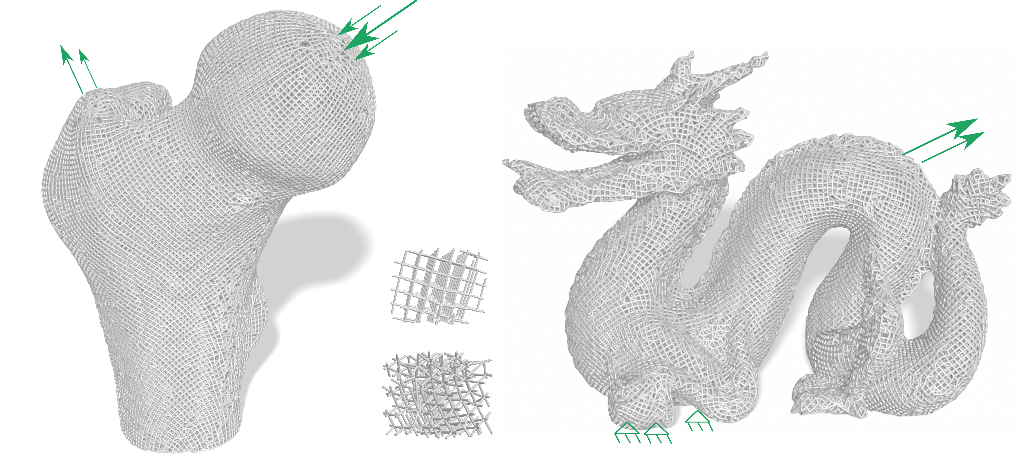}
\caption{Optimized 3D lattice structures for prescribed curved shapes. The optimized lattice structures possess spatial variations in orientation. The two samples are taken from inside the femur.
}\label{fig:lattice3Dfreeform}
\end{figure*}

\begin{figure*}[tb]
\centering
\includegraphics[width=1.0\linewidth]{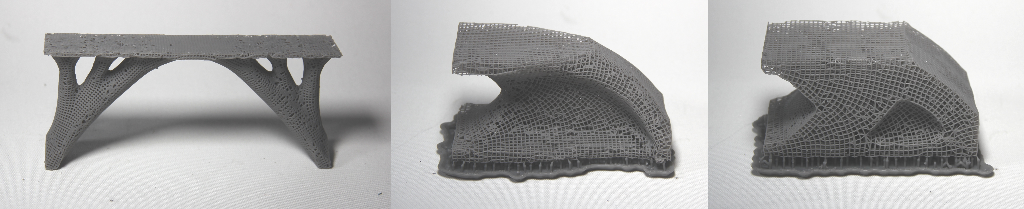}
\caption{Optimized lattice structures fabricated by 3D printing.
}\label{fig:prints1}
\end{figure*}

\begin{figure}[tb]
\centering
\includegraphics[width=\linewidth]{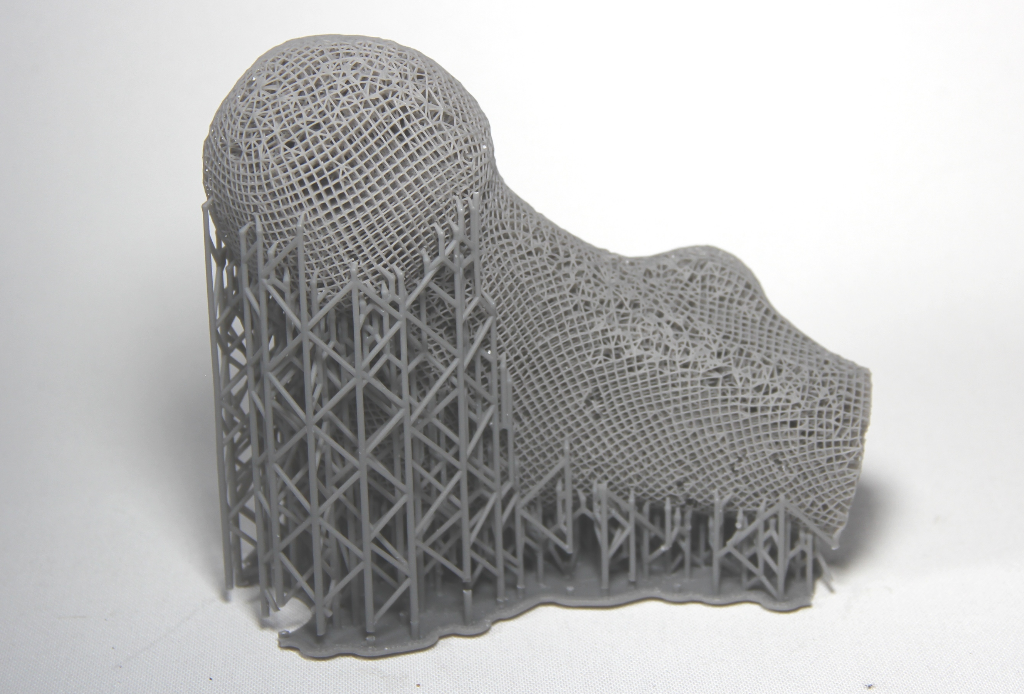}
\caption{3D printed femur with supports.
}\label{fig:prints-femur}
\vspace{-4mm}
\end{figure}

\begin{figure*}[!ht]
\centering
\def\svgwidth{\linewidth}
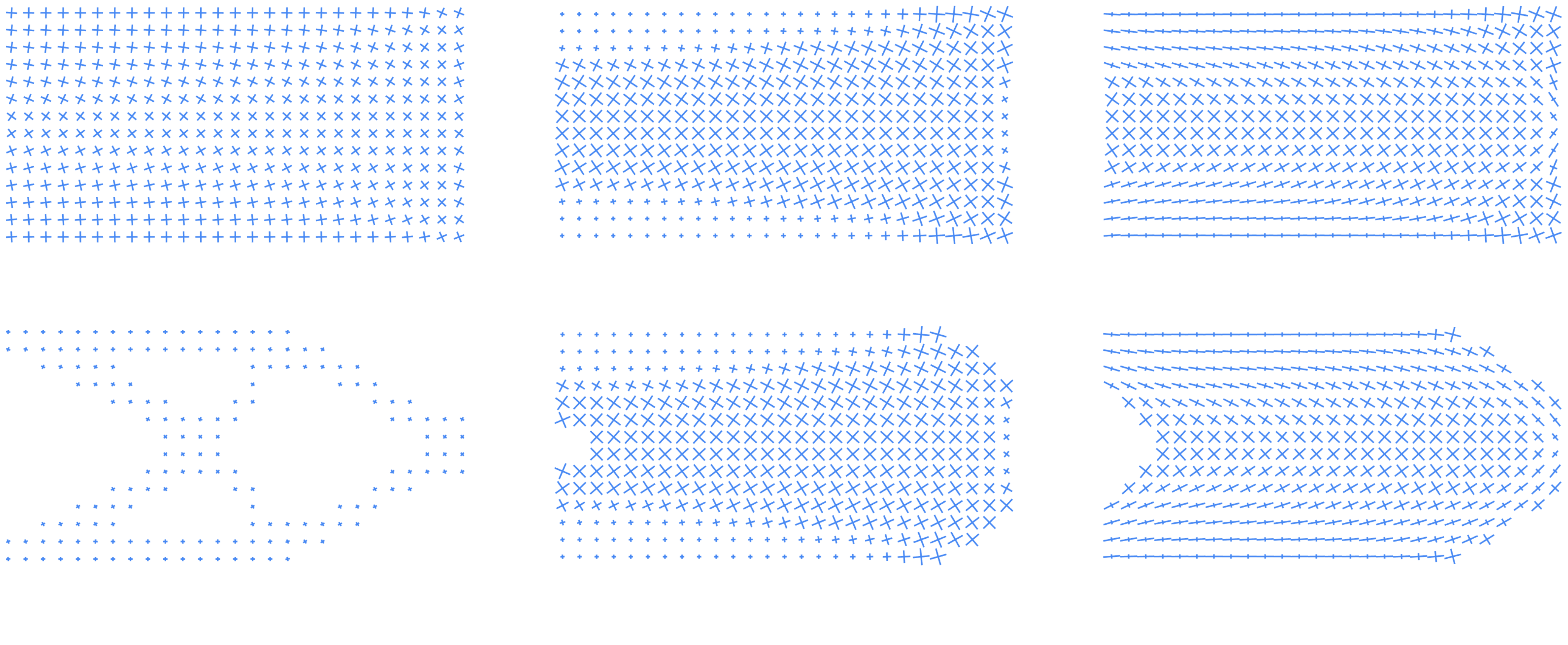
\caption{Visualization of optimized 2D fields corresponding to different design options.}
\label{fig:cantilever}
\vspace{4mm}
\centering
\includegraphics[width=0.98\linewidth]{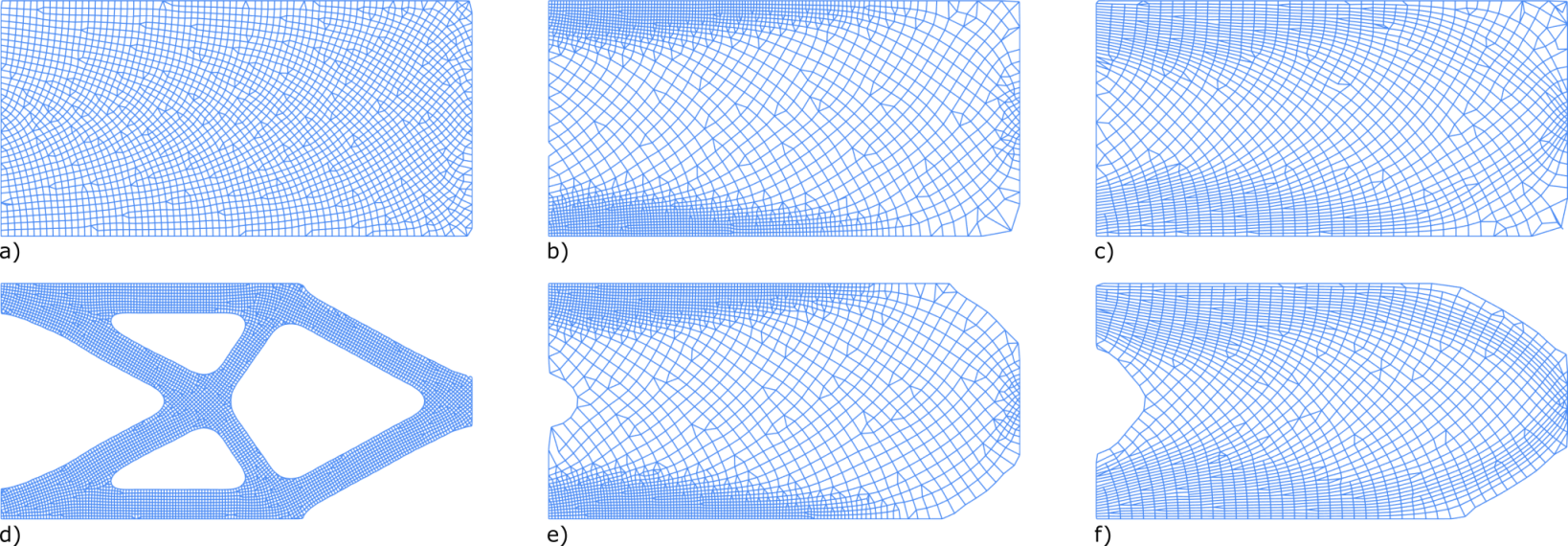}
\caption{Compiled lattice structures from the optimized, locally-defined lattice configuration (cf Fig.~\ref{fig:cantilever}).
}\label{fig:cantileverLattice}
\end{figure*}

\section{Results}
\label{sec:results}

\subsection{Examples}

Our method works for both 2D and 3D. 
Figure~\ref{fig:lattice2D} shows three optimized 2D lattice structures. In (a), the lattice distributes across the prescribed curved shape, with spatial variations in orientation, porosity, and anisotropy. In (b) and (c), the optimized lattices cover a subset of a rectangular design domain, with variations in orientation in (c) and additionally variations in porosity and anisotropy in (b). The unit cell in 2D is specified with $l=10t$, $\underline{\alpha}=1$, and $\overline{\alpha}=4$.

Figure~\ref{fig:lattice3D} show 3D lattice structures optimized by our method. Our method is also applicable to design lattices that spread over a prescribed 3D curved shape, as shown in Figure~\ref{fig:lattice3Dfreeform}. 3D examples are optimized with a unit cell using $l=4t$, $\underline{\alpha}=1$, and $\overline{\alpha}=2$. Figure~\ref{fig:prints1} shows fabricated bridge and cantilevers. The models are 3D printed with Form 2 which uses stereolithography (SLA). The dimension of models and the thickness of struts are scaled to comply with the volume and feature size of the printer. The printed femur (Figure~\ref{fig:prints-femur}) has a dimension of  $112.4\times77.9\times133.1\,mm^3$, with a thickness of $0.5\,mm$. The chair (Figure~\ref{fig:teaser}) is $110.8\times76.6\times142.1\,mm^3$, with a thickness of $0.6\,mm$. 

\subsection{Evaluation}

\subsubsection{Design options}
We evaluate the influence of various design options on the resulting lattice structures using the 2D cantilever ( Section~\ref{subsec:example}, Figure~\ref{fig:cantilever31}), with the fraction of solid material $\overline{v}=0.15$, and bounds for the scaling factors $\underline{\alpha}=1$ and $\overline{\alpha}=4$. The optimized fields and compiled lattice structures are shown in Figures~\ref{fig:cantilever} and~\ref{fig:cantileverLattice}, respectively. 

In the first row of Figure~\ref{fig:cantilever}, the fraction of lattice is fixed, $\varphi = 1$. Consequently the lattice distributes across the entire rectangular design domain. In (a) the scaling is also fixed, while in (b) the  optimization of scaling is enabled. The enlarged solution space leads a decrease in compliance (i.e. improved stiffness), $418.33$ (a) vs. $282.62$ (b). In (c), the scaling factors along individual axes are decoupled, resulting in a further decrease in compliance to $239.97$.

In the second row of Figure~\ref{fig:cantilever}, the fraction of lattice is optimized. Consequently, a shape evolves from the optimization, corresponding to $\varphi_e \ge 0.5$. Similar to the trend of compliance in the first row, it decreases from (d), to (e), and to (f), along with the increased flexibility in optimization. (f) has the smallest compliance among the six cases. It decreases from (a) by $44.39\%$. This confirms the significance of adaptive porosity and anisotropic features for stiffness maximization.

As a reference, an axis-aligned uniform lattice structure covering the entire domain (i.e. corresponding to the initialization of Figure~\ref{fig:cantilever}) is evaluated. Its compliance is $852.30$, which is more than twice larger than the design in Figure~\ref{fig:cantilever}a, and $3.66$ times larger than the design in Figure~\ref{fig:cantilever}f. This comparison confirms the importance of aligning anisotropic microstructures with internal stress directions for stiffness maximization.

\subsubsection{Accuracy}
To evaluate the accuracy of our lattice compilation method, we perform a comparison of the compliance predicted by homogenization with the compliance of lattice structures by a full-resolution finite element analysis. To this end, the six lattice structures in Figure~\ref{fig:cantileverLattice} are discretized by a finite element resolution of $4096\times2048$, and analyzed using a geometric multigrid elasticity solver~\cite{Amir2014SMO}. The comparison is summarized in Table~\ref{table:fullres}. The difference in compliance is between $2.89\%$ and $6.46\%$. This demonstrates that our lattice compilation introduces little error to the predicted performance from homogenization-based optimization. We note that homogenization theory assumes infinite periodicity of the cells, while for fabrication the compiled lattice has a finite physical size.

\begin{table}[h]
\caption{The difference in compliance predicted by homogenization and a full resolution analysis, for the lattice structures shown in Fig.~\ref{fig:cantileverLattice}.}
\label{table:fullres}
\begin{tabular}{l|cccccc}
          & a &   b     &   c     &    d    &    e    &    f    \\
          \hline
Homo.     & 418.33 & 282.62 & 239.97 & 332.81 & 277.27 & 232.64 \\
Full res. & 444.78 & 300.15 & 255.48 & 323.18 & 292.66 & 241.94 \\
Diff.     & 6.32\% & 6.20\% & 6.46\% & 2.89\% & 5.55\% & 4.00\%
\end{tabular}
\end{table}

\subsubsection{Computational performance}
Table~\ref{table:statistics} presents statistics of 3D model complexity and computational performance. The experiments were run on a standard desktop PC equipped with an Intel Xeon E5-1650 v3 processor (12 cores) running at 3.5 GHz, 64 GB RAM, and an Nvidia GTX1080 graphics card with 8 GB memory. The optimization and compilation together take less than 1 hour even for complex models such as the chair and femur. 

The group of columns 2-8 is related to lattice optimization. From the cantilever and bridge examples, it can be observed that with increasing design flexibility the compliance ($J_{com}$) decreases. This agrees with the 2D tests in Figure~\ref{fig:cantilever}. The increased design flexibility is also reflected by an increase of time associated with updating stiffness matrices, which is counted in $T_{FEA}$. The optimization time of the gradient-based solver for $\boldsymbol{\varphi}$ and $\boldsymbol{\alpha}$, $T_{Opt}$, increases accordingly as well.

The resolution of optimized fields is refined by a regular subdivision (1 element $\to$ $2^3$ elements), followed by tri-linear interpolation of the fields. While our lattice compilation algorithm takes a general graph as the input, in our implementation, we use triangle meshes and tetrahedral meshes which are purely for the convenience of computing vertex normal. This step costs 45$\sim$70 seconds (cantilevers, Figure~\ref{fig:lattice3D}) to 4 minutes and 26 seconds (chair, Figure~\ref{fig:teaser}). The refinement generates a large number of vertices (\#vertex) organized as tetrahedral elements (\#tet),  supplied to the lattice compilation. The compiled lattice has as many as 305k struts, for the femur model. Timings for pre-processing, i.e. building data structures ($T_{pre}$), local parameterization ($T_{posy}$), and graph extraction ($T_{extr}$) are reported.

\begin{figure*}[!ht]
\centering
\includegraphics[width=0.98\linewidth]{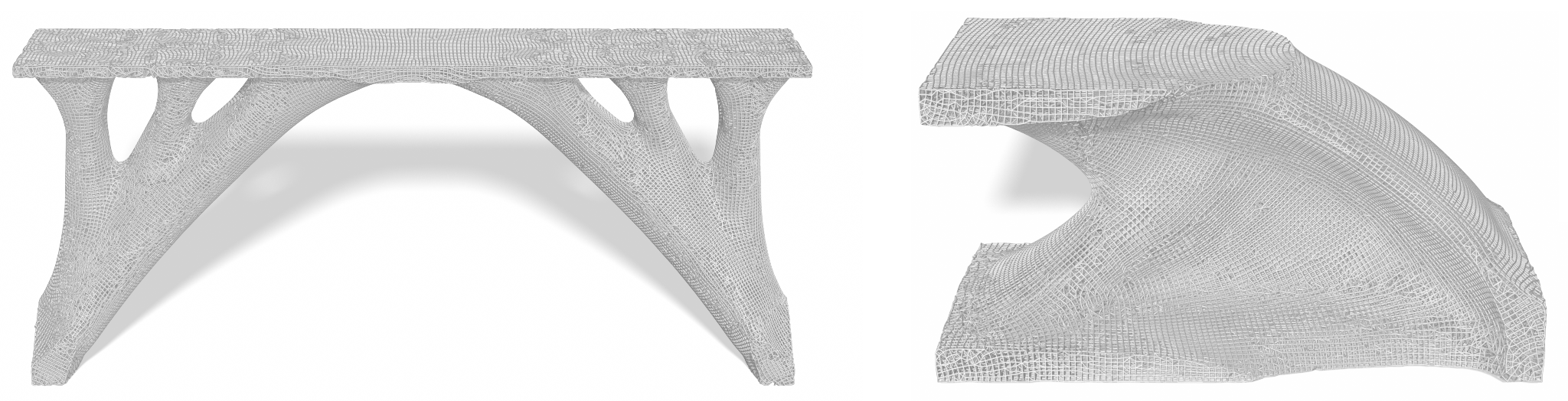}
\caption{Optimized lattice structures composed of 462k struts (bridge) and 351k struts (cantilever).}
\label{fig:refineTwice}
\vspace{-4mm}
\end{figure*}

In the last two rows, the optimized fields are refined twice (1 element $\to$ $4^3$ elements). This creates highly detailed lattice structures as shown in Figure~\ref{fig:refineTwice}.

\begin{table*}[h]
\footnotesize
\setlength{\tabcolsep}{3pt}
\caption{Statistics of 3D model complexity and computational performance. The timing is reported in minutes.}
\label{table:statistics}
\begin{tabular}{r || rrr | rr | rr || rr | r | rrr || r}
Model        & Resolution        & \#Ele. & Vol. & \#It. & $J_{com}$ & $T_{FEA}$      & $T_{Opt}$     & \#vertex & \#tet  & \#strut & $T_{pre}$       & $T_{posy}$       & $T_{extr}$        & $T_{Total}$     \\
\hline
3D cantilever (Fig.~\ref{fig:lattice3D}a)  & $100\times50\times50$   & 250k   & 0.2  & 60    & 110.84     & 3.11 & 0.85 & 0.89m    & 5.0m  & 48k     & 1.80 & 5.28 & 0.33     & 11.36\Tstrut\\
3D cantilever (Fig.~\ref{fig:lattice3D}b) & $100\times50\times50$   & 250k   & 0.2  & 60    & 96.03      & 3.97 & 1.62 & 1.65m    & 9.62m  & 87k   & 1.75 & 9.98 & 0.70 & 18.03\\
3D cantilever (Fig.~\ref{fig:lattice3D}c) & $100\times50\times50$   & 250k   & 0.2  & 60    & 85.85      & 5.83 & 2.65 & 1.62m    & 9.41m  & 25k     & 1.83 & 6.50 & 0.57 & 17.38\Bstrut\\
Bridge (Fig.~\ref{fig:lattice3D}d)      & $200\times38\times88$   & 644k   & 0.1  & 60    & 230.52     & 15.13 & 1.88 & 1.18m    & 6.54m  & 63k     & 1.41 & 7.09    & 0.47 & 25.97\Tstrut\\
Bridge (Fig.~\ref{fig:lattice3D}e)     & $200\times38\times88$   & 644k   & 0.1  & 60    & 177.86     & 16.76 & 3.80 & 2.02m    & 11.57m & 111k    & 2.84 & 13.58 & 1.14 & 38.12\\
Bridge (Fig.~\ref{fig:lattice3D}f)     & $200\times38\times88$   & 644k   & 0.1  & 60    & 149.96     & 21.10 & 6.29 & 1.89m    & 10.77m & 35k     & 2.40 & 8.29    & 0.72     & 38.79\Bstrut\\
chair (Fig.~\ref{fig:teaser})       & $140\times100\times200$ & 1.8m   & 0.1  & 60    & 193.5      & 30.92 & 5.03 & 3.32m    & 18.60m & 178k    & 4.15 & 18.66   & 1.87     & 60.63\Tstrut\Bstrut\\
femur (Fig.~\ref{fig:lattice3Dfreeform})       & $140\times93\times182$  & 696k   & 0.5  & 6     & 163.4      & 0.99 & 0       & 5.86m    & 14.26m & 305k    & 12.36 & 35.50 & 5.94 & 54.79\Tstrut\\
dragon (Fig.~\ref{fig:lattice3Dfreeform})      & $200\times90\times143$  & 461k   & 0.5  & 6     & 99.4       & 1.12 & 0       & 4.09m    & 23.31m & 200k    & 5.09    & 24.84 & 2.88    & 33.92\Bstrut\\
3D cantilever (Fig.~\ref{fig:refineTwice})  & $100\times50\times50$   & 250k   & 0.2  & 60    & 110.84     & 3.11 & 0.85 & 6.65m    & 38.50m  & 351k     & 7.19 & 33.44 & 6.25     & 50.84\Tstrut\\
Bridge (Fig.~\ref{fig:refineTwice})      & $200\times38\times88$   & 644k   & 0.1  & 60    & 230.52     & 15.13 & 1.88 & 8.64m    & 49.63m  & 462k     & 12.50 & 56.56    & 14.35 & 100.42\Bstrut\\
\end{tabular}
\end{table*}

\subsection{Comparison and Discussion}
\begin{figure*}[!ht]
\centering
\includegraphics[width=1.0\linewidth]{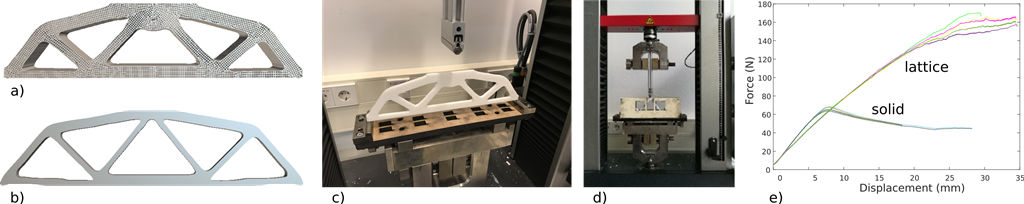}
\caption{Physical test comparing an optimize lattice structure (a) with an optimized solid structure (b). (c) and (d) show the experimental setup. From the force-displacement plots for multiple tests (e), it can be observed that, while the lattice structure is slightly less stiff, it supports a compressing force twice larger than the peak force supported by the solid counterpart.}
\label{fig:physicalTest}
\vspace{-4mm}
\end{figure*}

\noindent
\textbf{Comparison with solid structures from density-based topology optimization~\cite{Sigmund2001SMO}} \quad 
A 2D simply supported beam is optimized using our method and the classic density-based approach -- Solid Isotropic Material with Penalization (SIMP). The lattice and solid structure generated by our method and SIMP, are shown in Figure~\ref{fig:physicalTest} a) and b), respectively. The physical sizes are $294.8\times74.2\times60\;mm^3$, and the struts have an in-plane thickness of $0.8\;mm$, which is twice the nozzle size. They were fabricated by a Ultimaker S5 printer using flexible TPU material. While the digital models were designed using the same fraction of solid material, with 3D printing the lattice structure is heavier ($52$ gram vs. $46$ gram) due to the delicate tool-path.

The load condition of the 3D printed specimen is shown in Figure~\ref{fig:physicalTest}c. It is supported at the two ends on the bottom, while a downward force is applied on the top middle. To avoid out-of-plane buckling of these thin specimens, two wooden plates (with open square windows for observation) are placed to clamp the specimen (Figure~\ref{fig:physicalTest}d). Clamping plates are placed with a gap of $62\;mm$, slightly larger than the thickness of the specimen.

The force-displacement plots for multiple tests are shown in Figure~\ref{fig:physicalTest}e. The forces on the solid structure increase steeper than on the porous structure, meaning that the solid structure from SIMP has a higher stiffness. However, the forces on the solid structure turn down after they reach a peak of about $62$~N. This is due to the (in-plane) buckling of the compressed bars. In contrast, the lattice structure can support a maximum force that is twice larger before it buckles. This is due to the increased effective cross-section area of the substrctures. This test, in agreement with previous physical tests on 3D printed isotropic infill~\cite{Clausen2016Eng}, confirms the significance of lattice structures for buckling stability. We note that directly accounting for buckling stability in topology optimization is a much complicated problem, due to the less intuitive definition of the buckling mechanism and demanding eigenvalue problems~\cite{Ferrari2019SMO}. Optimizing structures with lattice materials provides an efficient and effective solution to increase buckling stability.



\begin{figure}[!ht]
\centering
\includegraphics[width=0.8\linewidth]{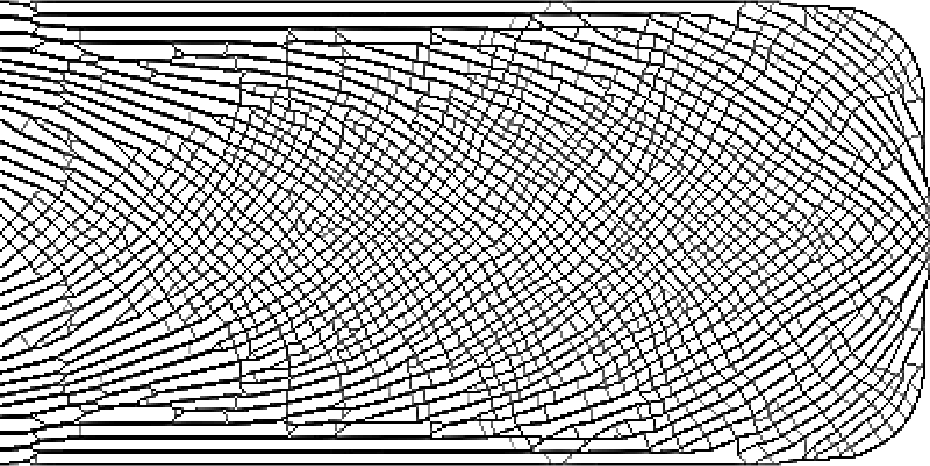}\\
\vspace{2mm}
\includegraphics[width=0.8\linewidth]{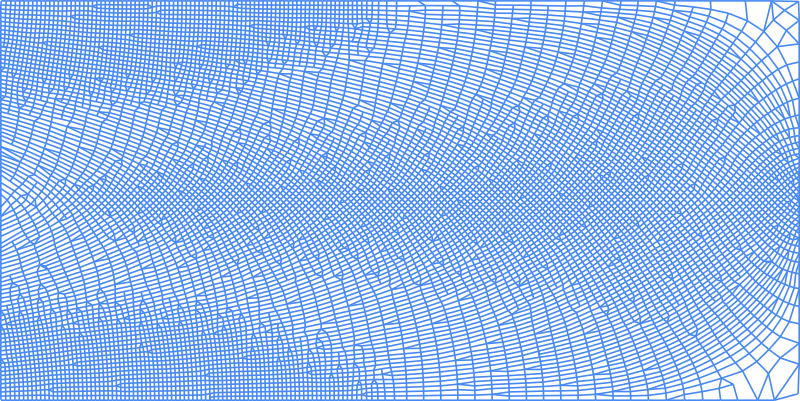}
\caption{Top: A bone-like porous structure generated by local volume constraints~\cite{Wu2018TVCG}. Bottom: A conforming lattice structure generated by the proposed method. The lattice is stiffer, with a compliance of $177.29$, compared to $184.64$ of the porous structure.}
\label{fig:compare-bone}
\end{figure}

\noindent
\textbf{Comparison with bone-like porous structures~\cite{Wu2018TVCG}} \quad 
Wu et al. proposed a density-based approach to design bone-like porous structures using constraints on local material volume~\cite{Wu2018TVCG}. Figure~\ref{fig:compare-bone} compares the porous structure and the lattice structure, generated with the same boundary conditions (see Fig.~\ref{fig:pipeline}a) and the same fraction of solid material. The porous structure was optimized with a local volume fraction of $0.36$, leading to a total volume fraction of $0.288$. We then run lattice optimization with this total volume fraction, with the design options of rotation and scaling. The bone-inspired infill was optimized with a finite element resolution of $400\times200$, while the lattice was obtained with a simulation resolution of $80\times40$. 

\begin{figure}[!ht]
\centering
\def\svgwidth{\linewidth}
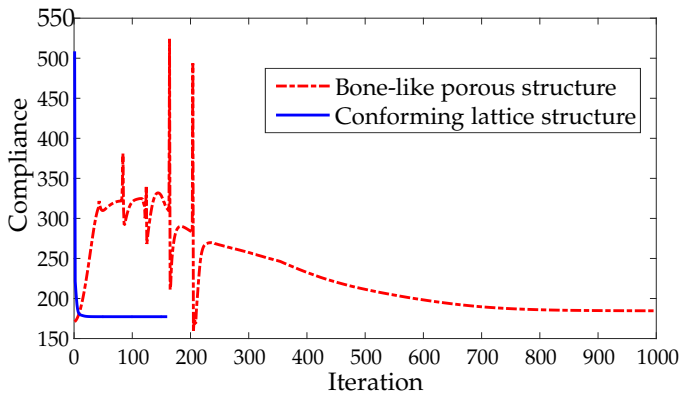
\caption{Convergence in compliance for the optimization of the bone-like porous structure and the conforming lattice structure.}
\label{fig:compare-convergence}
\vspace{-4mm}
\end{figure}

The convergence in compliance is plotted in Figure~\ref{fig:compare-convergence}. The compliance of bone-like infill and conforming lattice is $184.64$ and $177.29$, respectively, meaning that the lattice structure is stiffer. Lattice optimization converges much faster, and since it runs on a lower resolution, this leads to a significant speed up. The optimization of lattice took 1 minute and 7 seconds (60 iterations), while the optimization of bone-like infill cost 40 minutes (1000 iterations).

\noindent
\textbf{Discussion of Arora et al.~\cite{Arora2018:arxiv}} \quad 
Arora et al. proposed a method to construct a graph with its edges aligned with stress directions from simulation of the solid shape~\cite{Arora2018:arxiv}. This approach does not explicitly optimize the orientation, porosity, nor anisotropy. It relates to the option in our method with fixed $\varphi$ and $\alpha$ (cf. Figures~\ref{fig:cantilever}a and~\ref{fig:cantileverLattice}a). The result in Figure~\ref{fig:cantilever}f has demonstrated that with optimized porosity and anisotropy, the compliance reduces by $44.39\%$. We note that under the option of fixed $\varphi$ and $\alpha$, after aligning the lattice, our method re-calculates stress directions and update the lattice orientation. This leads a minor but noticeable decrease in compliance than aligning the lattice with stress directions from the solid shape (420.58 $\to$ 418.90). 

Our lattice compilation approach is scalable, for example, the number of struts is more than two orders of magnitude compared with examples shown in~\cite{Arora2018:arxiv}. This allows to design highly detailed lattice structures. Figure~\ref{fig:refineTwice} shows optimized lattice structures with 462k struts (bridge) and 351k struts (cantilever). 
\section{Conclusions}
\label{sec:conclusions}
In this paper we have presented a novel method to design conforming lattice structures by extending homogenization-based topology optimization and field-aligned parameterization. It can compute not only an optimized lattice structure that occupies certain subregions of regular design domains but also lattices that spread over prescribed (curved) shapes. The optimized lattice structures conform with principal stress directions and the boundary of the (optimized) shape. Our method is scalable and allows to optimize highly detailed lattice structures, which can be fabricated by 3D printing. Numerical analysis on different design options confirms the importance of aligning anisotropic lattice with internal stress directions and the importance of lattice gradation in porosity and anisotropy. The compiled lattice structure, by full resolution finite element analysis, has a compliance very close to the compliance predicted by homogenization-based optimization. By physical tests we demonstrate that the optimized lattice structure can support a buckling load twice as large as topology optimized solid structures, at the price of a slight decrease in stiffness. Besides quantified structural performance, the optimized conforming lattice structures look remarkably appealing. 

\noindent
\textbf{Future work} \quad Our method generates lattice structures particularly optimized for mechanical properties. It provides options to steer the optimization by configuring the design variables, and to adapt the output graph resolution in lattice compilation. It is found that the generated lattice, in certain areas where the stress is small, is less regular. This is attributed to the rapid spatial variation in the underlying stress directions. In some applications the designer might wish to trade the mechanical performance for regularity also in such rapidly change areas. An elegant solution for this balancing, ideally embedded in the lattice optimization step, remains an open question. 

\section*{Acknowledgements}
The authors gratefully acknowledge the support from the LEaDing Fellows Programme at the Delft University of Technology, which has received funding from the European Union's Horizon 2020 research and innovation programme under the Marie Skłodowska-Curie grant agreement No.~707404. Weiming Wang wishes to thank the Natural Science Foundation of China (No.~61702079).

\appendix
\section{Appendix}
\label{sec:appendix}

Denoting a $3 \times 3$ rotation matrix by
\begin{equation}
R =  \begin{pmatrix}
 l_1 & l_2 & l_3 \\
 m_1 & m_2 & m_3 \\
 n_1 & n_2 & n_3
 \end{pmatrix},
\end{equation}
the $6 \times 6$ rotation matrix for the elasticity tensor in engineering notation is written as
\begin{equation}
\overline{R} = 
 \begin{pmatrix}
  A & B \\
  C & D
 \end{pmatrix},
\end{equation}
with 
\begin{equation}
{A} = 
 \begin{pmatrix}
  l_1^2 & m_1^2 & n_1^2 \\
  l_2^2 & m_2^2 & n_2^2 \\
  l_3^2 & m_3^2 & n_3^2 
 \end{pmatrix}, \\
\end{equation}

\begin{equation}
{B} = 
 \begin{pmatrix}
  2m_1 n_1 & 2n_1 l_1 & 2l_1 m_1 \\
  2m_2 n_2 & 2n_2 l_2 & 2l_2 m_2 \\
  2m_3 n_3 & 2n_3 l_3 & 2l_3 m_3
 \end{pmatrix}, \\
\end{equation}

\begin{equation}
{C} = 
 \begin{pmatrix}
  l_2 l_3 & m_2 m_3 & n_2 n_3 \\
  l_3 l_1 & m_3 m_1 & n_3 n_1 \\
  l_1 l_2 & m_1 m_2 & n_1 n_2
 \end{pmatrix},
\end{equation}
and
\begin{equation}
{D} = 
 \begin{pmatrix}
  m_2 n_3 + m_3 n_2 & n_2 l_3 + n_3 l_2 & m_2 l_3 + m_3 l_2 \\
  m_3 n_1 + m_1 n_3 & n_3 l_1 + n_1 l_3 & m_3 l_1 + m_1 l_3 \\
  m_1 n_2 + m_2 n_1 & n_1 l_2 + n_2 l_1 & m_1 l_2 + m_2 l_1
 \end{pmatrix}.
\end{equation}


\bibliographystyle{IEEEtran_noURL}
\bibliography{_bibliography,meshing,fields}

\begin{IEEEbiography}[{\includegraphics[width=1in,height=1.25in,clip,keepaspectratio]{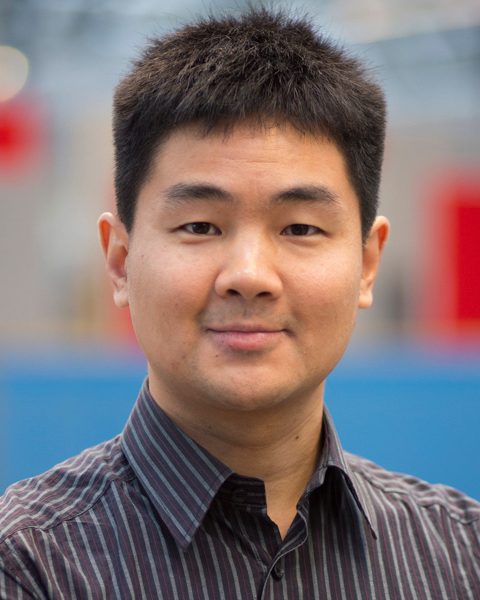}}]{Dr. Jun Wu} is an assistant professor at the Department of Design Engineering, Delft University of Technology, the Netherlands. Before this, he was a Marie Curie postdoc fellow at the Department of Mechanical Engineering, Technical University of Denmark. He obtained a PhD in Computer Science in 2015 from TU Munich, Germany, and a PhD in Mechanical Engineering in 2012 from Beihang University, Beijing, China. His research is focused on computational design and digital fabrication, with an emphasis on topology optimization.
\end{IEEEbiography}

\begin{IEEEbiography}[{\includegraphics[width=1in,height=1.25in,clip,keepaspectratio]{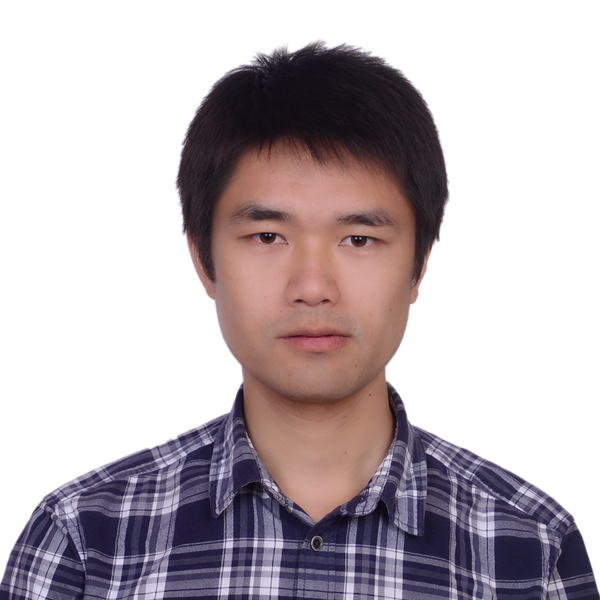}}]{Dr. Weiming Wang} is a Marie Curie postdoc fellow at the Department of Design Engineering, Delft University of Technology, the Netherlands, and a lecturer at the School of Mathematical Sciences, Dalian University of Technology, China. He obtained his PhD in the School of Mathematical Sciences in 2016 from Dalian University of Technology, China. His research is focused on digital fabrication and geometry processing.
\end{IEEEbiography}

\begin{IEEEbiography}[{\includegraphics[width=1in,height=1.25in,clip,keepaspectratio]{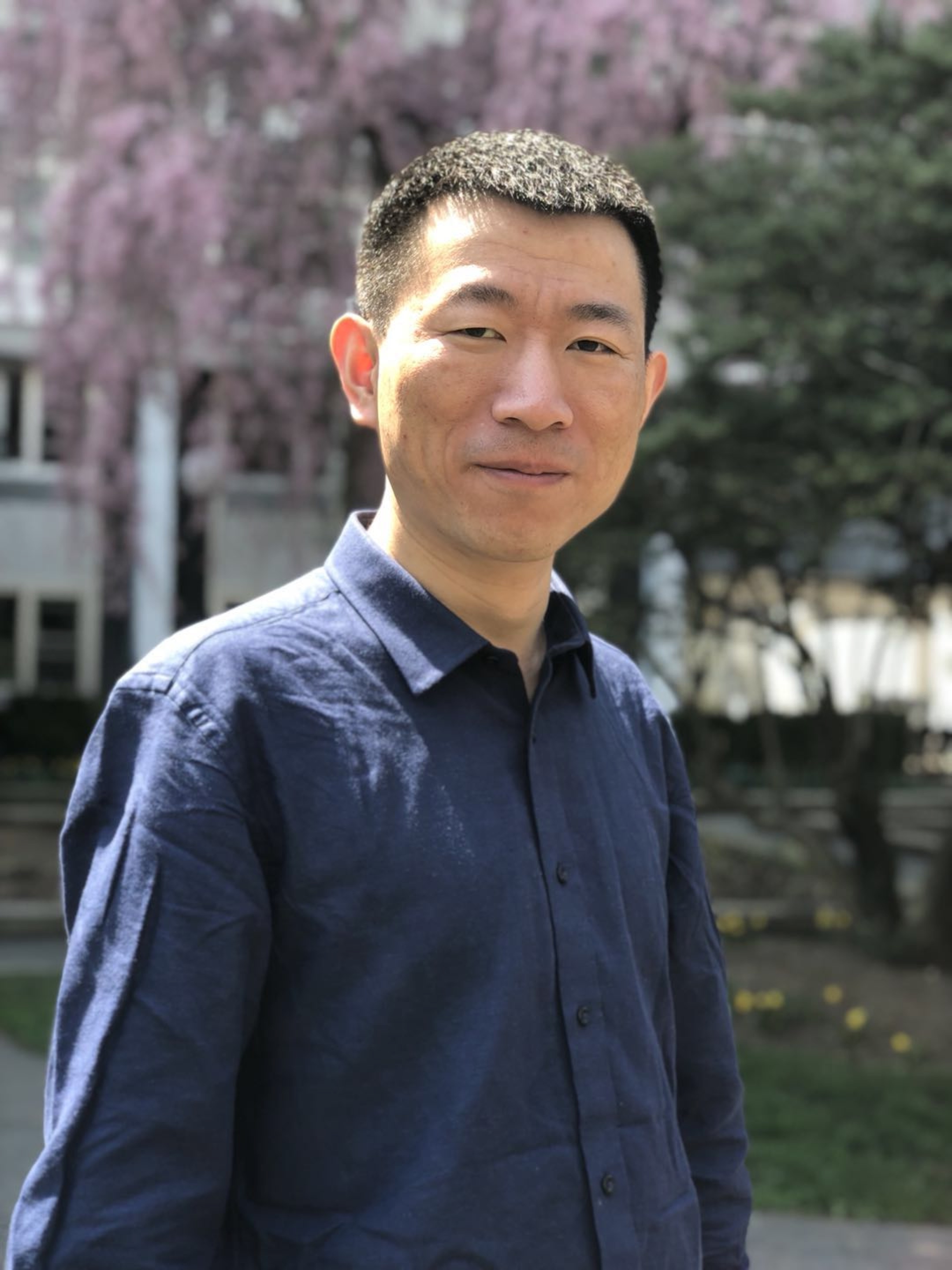}}]{Dr. Xifeng Gao} is an assistant professor of the Computer Science Department at Florida State University. Dr. Gao was a PostDoc for two years at the Courant Institute of Mathematical Sciences of New York University. He received his Ph.D. degree in 2016 and won the best Ph.D. dissertation award from the Department of Computer Science at the University of Houston. Dr. Gao has wide research interests that are related to geometry, such as Computer Graphics, Visualization, Multimedia Processing, Robotics, and Digital Fabrication.
\end{IEEEbiography}

\end{document}